\documentclass[a4paper]{article}

\usepackage[english]{babel}
\usepackage[utf8]{inputenc}
\usepackage{amsmath}
\usepackage[colorinlistoftodos]{todonotes}
\usepackage[margin=1.25in]{geometry}
\usepackage[raggedright]{titlesec}

\usepackage{authblk}
\title{Ultrafast-laser-absorption spectroscopy in the mid-infrared for single-shot, calibration-free temperature and species measurements in low- and high-pressure combustion gases}
\author[1,*]{Ryan J. Tancin}
\author[1,$\dagger$]{Christopher S. Goldenstein}
\affil[1]{School of Mechanical Engineering, Purdue University, West Lafayette, IN 47907, USA}
\affil[*]{rjtancin@gmail.com}
\affil[$\dagger$]{csgoldenstein@purdue.edu}
\begin{document}
\maketitle





\begin{abstract}
This manuscript presents an ultrafast-laser-absorption-spectroscopy (ULAS) diagnostic capable of providing calibration-free, single-shot measurements of temperature and CO at 5 kHz in combustion gases at low and high pressures. A detailed description of the spectral-fitting routine, data-processing procedures, determination of the instrument response function, and practical considerations for imaging ultrashort pulses in the mid-infrared are presented. The accuracy of the diagnostic was validated at 1000 K and pressures up to 40 bar in a heated-gas cell before being applied to characterize the spatiotemporal evolution of temperature and CO in AP-HTPB and AP-HTPB-aluminum propellant flames at pressures between 1 and 40 bar. The results presented here demonstrate that ULAS in the mid-IR can provide high-fidelity, calibration-free measurements of gas properties with sub-nanosecond time resolution in harsh, high-pressure combustion environments representative of rocket motors.
\end{abstract}
\section{Introduction}

Laser-absorption spectroscopy (LAS) is widely used to provide non-invasive measurements of temperature, chemical species, pressure, and velocity. The demand for LAS measurements in high pressure and highly transient combustion and propulsion environments (e.g., detonation engines, rocket motors)
continues to propel innovations in LAS diagnostics \cite{nair2020mhz, tancin2020ultrafast, loparo2020broadband, Draper2019, makowiecki2021mid} and data-processing techniques \cite{goldenstein2017infrared, cole2019baseline, goldenstein2020cepstral}. Several of these innovations have focused on increasing the usable spectral bandwidth of LAS diagnostics while maintaining or improving temporal resolution. 

LAS diagnostics utilizing high spectral bandwidth benefit from the greater diversity of spectral information collected. This helps mitigate some challenges associated with determining the non-absorbing baseline and reduced levels of differential absorption at elevated pressures \cite{goldenstein2017infrared,tancin2021ultrafast,Radhakrishna2021}, as well as improve the dynamic-range of thermometry \cite{An2011}. Broadband LAS diagnostics must, however, achieve fast temporal resolution in order to characterize highly transient processes such as chemical or internal non-equilibrium and to avoid time-varying, non-absorbing transmission losses such as those from beam steering or particle scattering.
To date, broadband LAS diagnostics with fast time resolution can be grouped into two classes: (1) those employing broadly tunable CW light and (2) those employing ultrashort pulses. Regarding the former, vertical cavity surface-emitting lasers (VCSELs) \cite{sanders2002wavelength,rein2017measurements} and rapidly tunable external-cavity quantum-cascade lasers (EC-QCLs) \cite{Strand2019measurement} have been utilized in the near-IR and mid-IR, respectively, to provide broadband (10s to 100s of cm$^{-1}$) measurements of atomic and molecular absorbance spectra in combustion environments; however, such light sources are currently available at only a few central wavelengths. As a result, here we will focus on diagnostics employing broadband pulses within various architectures, namely, supercontinuum lasers (SCLs) \cite{sanders2002,werblinski2017high}, dual-frequency-comb spectrometers (DFCS)  \cite{makowiecki2021mid,draper2019broadband}, optical parametric oscillators (OPOs), \cite{loparo2020broadband} and optical parametric amplifiers (OPAs) \cite{tancin2020ultrafast,stauffer2021broadband,mcgrane2004shock, zamkov2007ultrafast}. 

Regarding SCLs, Sanders \cite{sanders2002} utilized a femtosecond-laser-pumped SCL producing 1000 cm$^{-1}$ of bandwidth near 1.56 $\mu$m to acquire temperature and species measurements at 50 kHz in gases at 10 bar. Similarly, Werblinski et al. \cite{werblinski2017high} used an SCL with 500 cm$^{-1}$ of bandwidth near 1.4 $\mu$m to acquire 10 kHz measurements of temperature, pressure and H$_2$O in a rapid compression machine (RCM) at pressures up to 65 bar. While the large bandwidth and repetition rate of such diagnostics is extremely attractive, SCL-based diagnostics are often limited by pronounced shot-to-shot intensity variations (requiring multi-shot averaging) and complexities associated with accurately modeling various instrument-broadening sources \cite{emmert2018data}, particularly when dispersing fibers are used. More recently, DFCS has emerged as an effective tool for providing high-fidelity, broadband measurements of absorbance spectra with unprecedented spectral resolution. For example, Draper et al. \cite{draper2019broadband} utilized a DFCS with 160 cm$^{-1}$ of bandwidth near 1660 nm to acquire 1.4 kHz measurements of temperature and CH$_4$ at up to 30 bar inside of an RCM. However, DFCS also typically requires time-averaging to achieve high signal-to-noise ratio (SNR) and such diagnostics in the mid-IR are still emerging \cite{makowiecki2021mid,pinkowski2020quantum}. 

Last, the recent advent of high-speed, cryogenically cooled, mid-IR cameras has, in part, enabled the first single-shot, ultrafast (i.e., sub-nanosecond) absorption measurements in the mid-IR \cite{tancin2020ultrafast}, as well as broadband absorption measurements with near-microsecond time resolution at 10s of kHz \cite{loparo2020broadband}. Regarding the latter, Loparo et al. \cite{loparo2020broadband} employed a mid-IR OPO which produced ultrashort pulses at 100 MHz to provide 476 cm$^{-1}$ of bandwidth near 3.25 $\mu$m. This enabled the authors to acquire measurements of CH$_4$, C$_2$H$_4$ and C$_2$H$_6$ at 21.5 kHz in a shock tube at pressures near 2.5 bar. To date, this technique has demonstrated a measurement time resolution of 37 $\mu$s (set by the integration time of the camera).

Here we build on our recent development of ultrafast laser-absorption spectroscopy (ULAS) \cite{tancin2020ultrafast} to provide the first single-shot, ultrafast, mid-IR measurements of temperature and species in high-pressure and multi-phase combustion gases. This was made possible through advancements in the optical setup employed and several notable refinements to the spectral-modeling and \mbox{-fitting} procedures. For the first time, this manuscript describes these advancements and the key principles of ULAS in detail, while also providing (1) a detailed discussion of practical considerations for imaging ultrashort pulses in the mid-IR and (2) demonstrating a proof-of-concept for 1D resolved, intra-pulse ULAS measurements of gas properties.

\section{Experimental setup}
\subsection{Optical setup}
\label{sect: optical setup}

\begin{figure}[b!]
\centering
\includegraphics[width=1\textwidth]{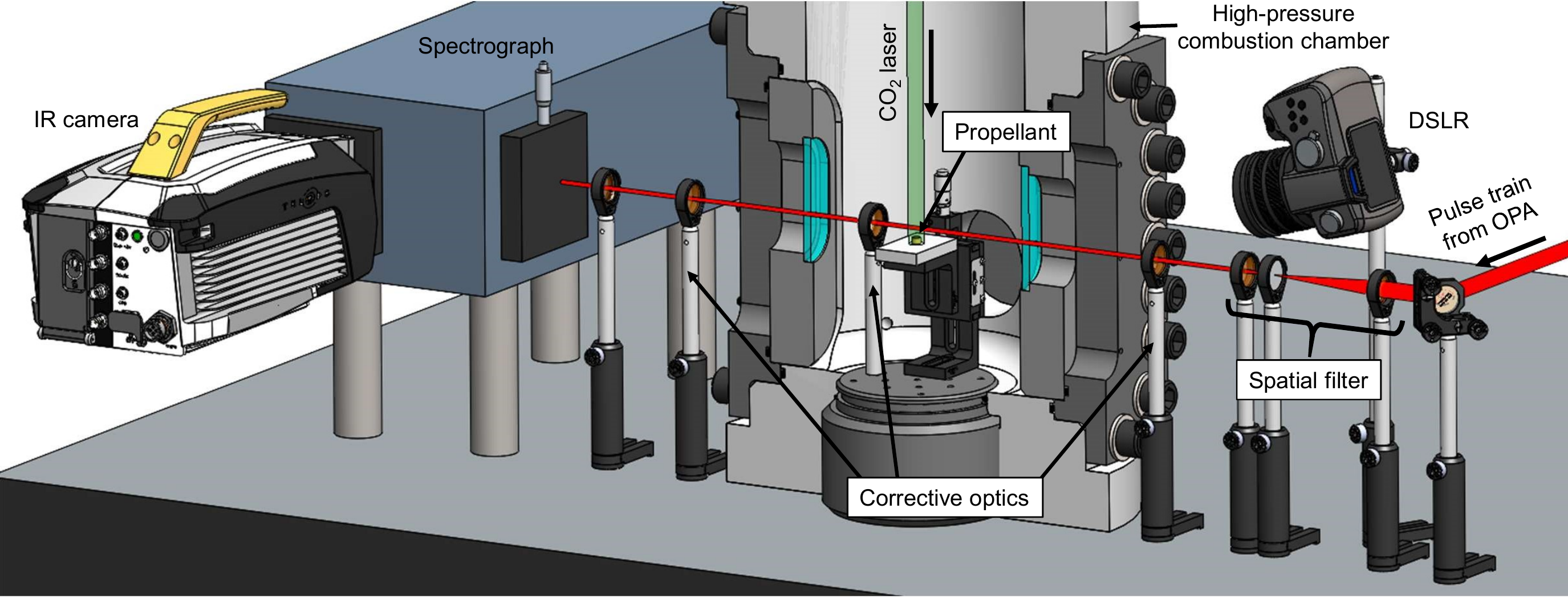}
\caption{Schematic illustrating the optical setup used to characterize propellant flames with ULAS.}
\label{fig:setup}
\end{figure}

Figure \ref{fig:setup} shows a schematic of the optical setup used for this work. For brevity, only the most pertinent details and recent changes will be discussed here. The ultrafast-laser system (described in previous work \cite{tancin2020ultrafast,tancin2021ultrafast,Radhakrishna2021,tancin2020ultrafast_AIAA}) consists of a Ti:Sapphire oscillator which produces 55 femtosecond pulses centered near 800 nm at 80 MHz. The pulses were selectively amplified at 5 kHz and subsequently converted into the mid-IR using  optical parametric amplification (OPA) and difference-frequency generation to achieve a center wavelength near 4.9 $\mu$m.

After wavelength conversion, a MgF$_2$ Rochon-prism polarizer was used to attenuate the pulse energy to $\approx$ 2 $\mu$J. Next, a spatial filter was used to improve beam quality and reduce the $1/e^2$ beam diameter to $\approx$ 2 mm. The laser beam was then directed through the test gas within either a heated-gas cell or propellant flame. Corrective optics (discussed at length in \cite{tancin2021ultrafast}) were applied here to reduce the impact of beam steering on flame measurements. These optics consisted of a lens telescope ($f$ = 200 mm and $f$ = 25.4 mm) with the focal point centered at the propellant flame and a final lens ($f$ = 250 mm) near the spectrograph (see Fig. \ref{fig:setup}). The former acted to partially correct any angular deflection caused by beam steering, minimizing the loss of signal, while the latter aids in boosting signal counts by reducing the beam size incident on the focal plane array (FPA) of the IR camera. The transmitted pulses were then focused onto the input slit of a spectrograph (Andor Shamrock 500i) using a CaF$_2$, plano-convex, cylindrical lens with a focal length of 75 mm. The spectrograph employed a reflective diffraction grating to disperse the light onto the InSb FPA of a high-speed, infrared camera (Telops FAST-IR 2K).  In propellant tests, images of single-shot spectra were averaged across the spatial dimension of the image to improve the measurement SNR. A high-resolution (300 lines/mm) diffraction grating providing 10 cm$^{-1}$ of bandwidth (near 5 $\mu$m) and 0.3 nm resolution was used for propellant tests at atmospheric pressure. A low resolution (150 lines/mm) grating providing 35 cm$^{-1}$ of bandwidth and 0.6 nm resolution was used for high-pressure propellant tests. Both gratings were used at all pressures for gas cell tests.

\subsection{Gas-cell experiments}
\label{sect: gas-cell setup}

ULAS measurements were acquired in a high-uniformity, heated, static-gas cell (described by Schwarm et al. \cite{schwarm2019high}) to validate the accuracy of the diagnostic at high temperature and pressure, as well as to determine the instrument response function (IRF) via a multi-spectrum fitting routine (MSFR) discussed in detail in Section \ref{sect: MSFR}. Spectra were measured in a mixture of 2\% CO and 1.8\% CO$_2$ by mole with a balance of nitrogen at 1000 $\pm$ 20 K and pressures of 0.3, 1, 3, 10, 20, and 40 bar. The gas pressure was measured using a UNIK-5000 pressure transducer with a full-scale range of 0 to 70 bar and an accuracy of $\pm$ 0.028 bar. Temperature was measured using three type-K thermocouples with a nominal accuracy of 0.75\% (at $\geq$ 600 K) fixed to the outside of the gas cell as in \cite{schwarm2019high}. The gas cell was equipped with 15 cm long, wedged, CaF$_2$ rods which provided an absorbing path length of 9.4 cm at room temperature. 

\subsection{Laser-ignited propellant flames}
\label{sect: propellant setup}

Cylindrical propellant strands with a height of 4 to 6 mm and a diameter of 7 mm were tested. The particle size of ammonium perchlorate (AP) was distributed bimodally with mean diameters of 200 $\mu$m and 20 $\mu$m in a coarse-to-fine ratio of 4:1. On a mass basis, the binder consisted of 76.33\% hydroxyl-terminated polybutadiene (HTPB) with 15.05\% isodecyl pelargonate as the plasticizer, and 8.62\% modified MDI isocyanate as the curative. In the aluminized propellants, Valimet H-30 aluminum with a mean particle diameter of 31 $\mu$m was used. The AP-HTPB propellant consisted of 80\% AP by mass, whereas the AP-HTPB-Al propellant was 68\% AP, 15\% aluminum and 17\% binder by mass. The propellant was cast into plastic tubes and cured for a minimum of 3 days at room temperature.

\begin{figure}[hbt!]
\centering
\includegraphics[width=0.4\textwidth]{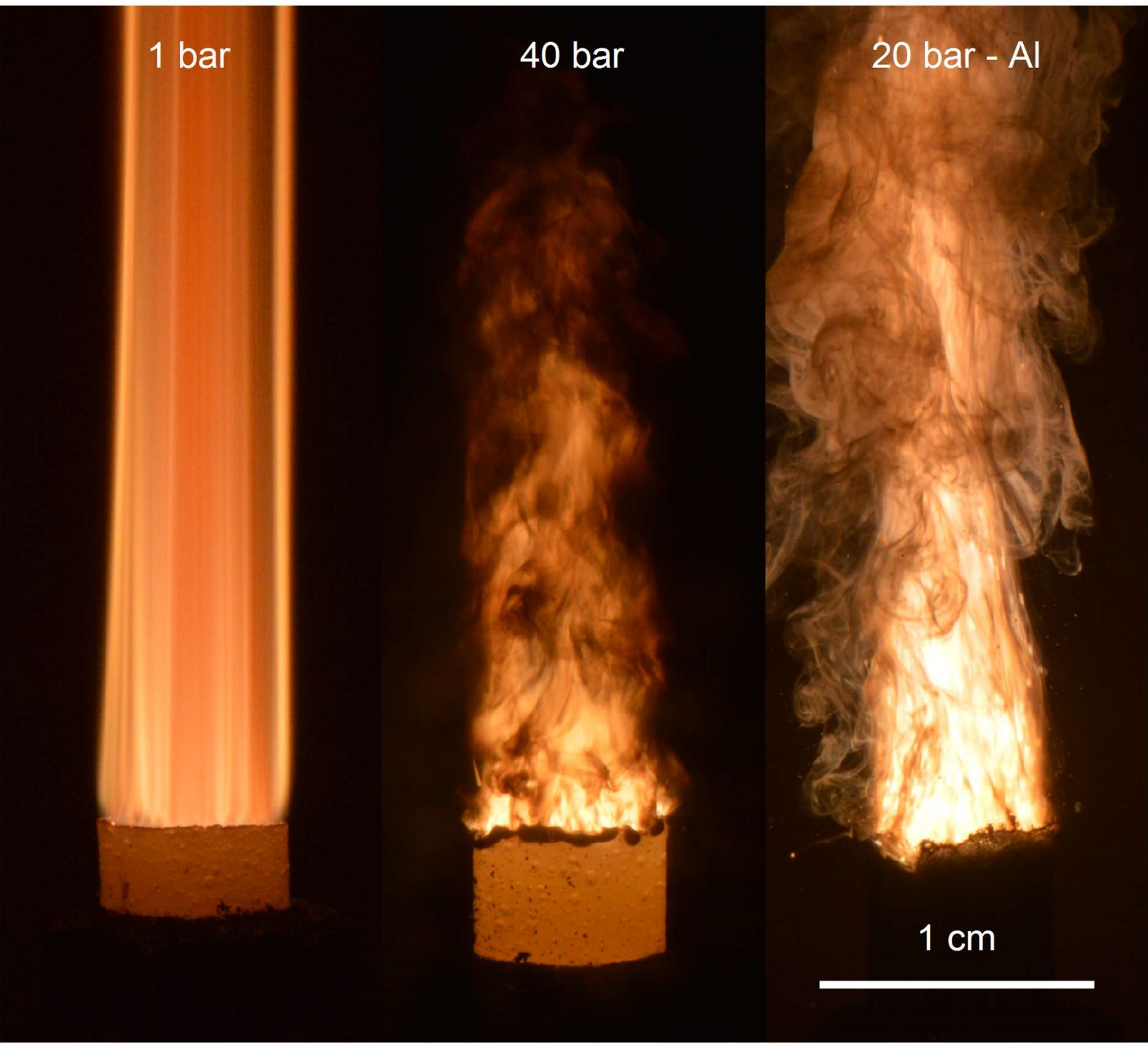}
\caption{DSLR images of laser-ignited AP-HTPB propellant flames at atmospheric- and high-pressure, without aluminum (left two panels) and with aluminum (right panel).}
\label{fig:flame_pics}
\end{figure}

The propellant strands were ignited using a CO$_2$ laser providing an optical intensity of 38 W/cm$^2$ to the surface of the strand. The data acquisition system was triggered 100 ms prior to the CO$_2$ laser turning on to allow for the baseline laser intensity ($I_0$) to be recorded. The CO$_2$ laser was turned on for 200 ms (400 ms at 1 bar) to ignite the strand. Video of each test was recorded with a Nikon D3200 DSLR camera to ensure that ignition and burning of each strand was uniform, and to provide an estimate of the absorbing path-length. The propellant flames reached a quasi-steady state approximately 100 ms after ignition. Figure \ref{fig:flame_pics} shows images of the propellant flames acquired using the DSLR camera. 

All propellant tests were conducted in a high-pressure combustion chamber (HPCC), described in detail by Tancin et al. \cite{Tancin2019_HPCC}. The ULAS beam was aligned through the centerline of the propellant for all tests. Measurements were acquired at a height of 2 mm and 1 cm relative to the initial surface location of the propellant strand for non-aluminized and aluminized propellants, respectively. Prior to each test, the chamber was evacuated using a vacuum roughing pump before being pressurized with an argon bath gas. A time history of pressure was recorded during each test enabling the data-processing routine to use an updated pressure for each subsequent laser shot. The gas pressure increased by $\approx$ 0.6 bar during each test. For tests at atmospheric pressure, the HPCC vent valve was partially opened immediately prior to ignition, such that no pressure rise occurred during the test.

\section{High-fidelity modeling and spectral-fitting of ULAS spectra}
\label{sect: fitting routine}

\begin{figure}[hbt!]
\centering
\includegraphics[width=0.97\textwidth]{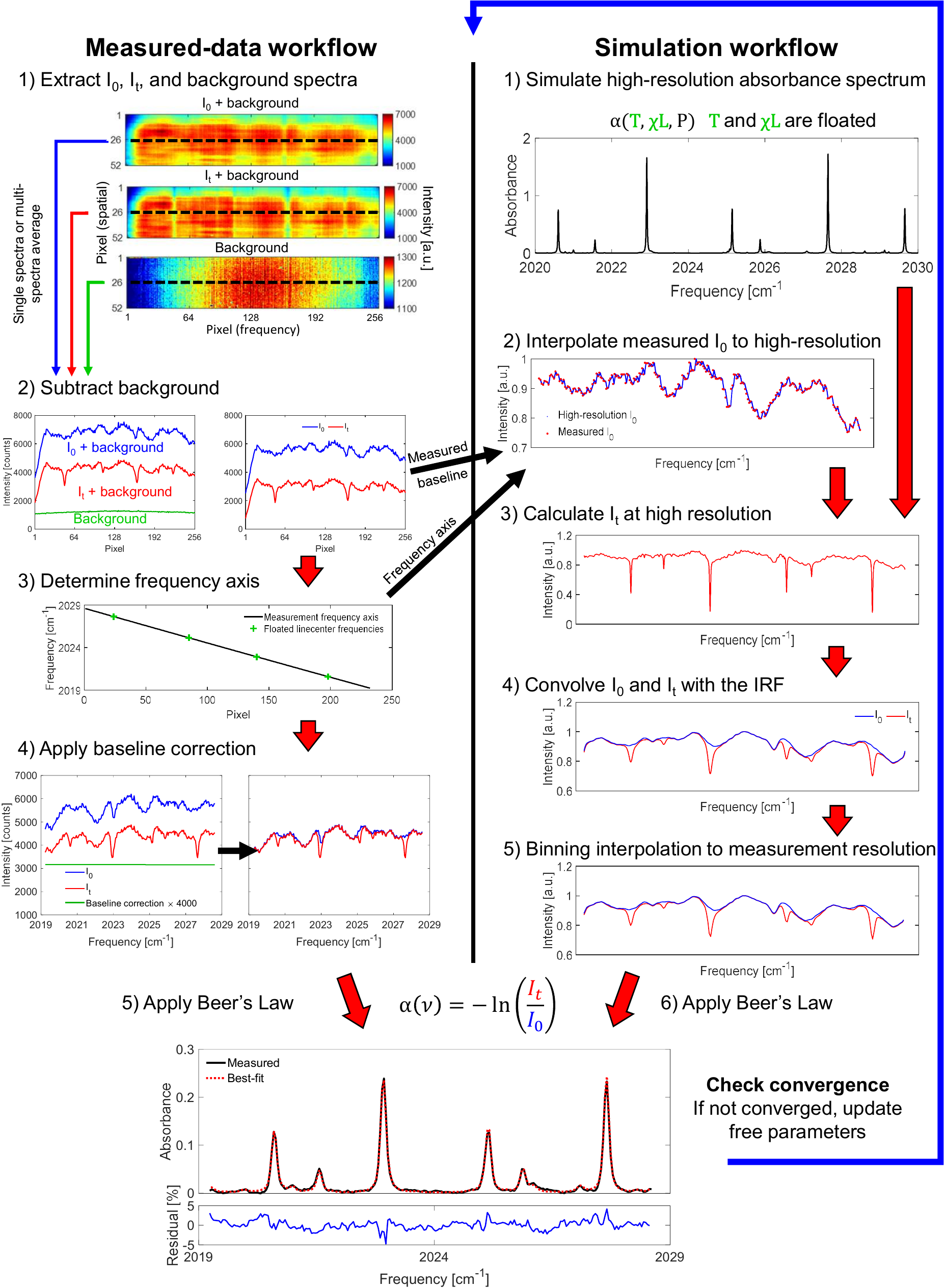}
\caption{Flowchart for the spectral-fitting routine used to infer gas properties from ULAS measurements. 
}
\label{fig:flowchart}
\end{figure}

This section provides a detailed description of the data-processing and spectroscopic modeling procedures used to infer gas properties from ULAS measurements of transmitted intensity spectra. In essence, a nonlinear least-squares fitting routine is used to fit simulated absorbance spectra to measured absorbance spectra. There are two workflows: (1) processing measured data and (2) simulating absorbance spectra which are described below and illustrated in Fig. \ref{fig:flowchart}.

The procedure for processing measured spectra consists of five key steps.
\begin{enumerate}
    \item Acquire images of $I_0$, transmitted laser intensity ($I_t$) and background emission.
    
    \item Subtract the appropriate background from images of $I_0$ and $I_t$. In propellant-flame experiments, the time-varying background was measured between successive laser-shots by acquiring images at twice the repetition rate of the laser.

    \item Assign a frequency axis to the measured spectra. Spectra are extracted from images of $I_0$ and $I_t$ in the form of a vector. The indices corresponding to 4 or 5 prominent spectral features (e.g., the linecenter of a strong transition) are assigned frequencies which are free parameters in the spectral-fitting routine. These ``control points'' allow a complete frequency axis to be mapped to the measured spectrum through cubic-spline interpolation. Cubic-spline interpolation is preferred over a linear or higher-order polynomial interpolation because the former doesn't account for slight nonlinearities in the frequency axis and the latter often leads to instability in the fitting routine.

    \item Adjust the baseline laser intensity in order to account for shot-to-shot laser fluctuations and non-absorbing transmission losses. These fluctuations exhibit a weak wavelength dependence and are accounted for by multiplying $I_0$ by a linear correction factor. Higher order corrections may yield slightly smaller residuals, but risk coupling with other free parameters and introducing error into the measurements \cite{goldenstein2020cepstral}.
    
    \item Apply Beer's Law to calculate the measured absorbance spectrum ($\alpha_{meas}$).
    
\end{enumerate}

The simulation-side workflow consists of six key steps described below and in Fig. \ref{fig:setup} (right). These steps were chosen to faithfully capture the physics which govern the optical setup and data-acquisition system. Subscripts ``$HR$'' and ``$conv$'' indicate high spectral-resolution and post-convolution with the IRF, respectively.

\begin{enumerate}
    \item Simulate an absorbance spectrum ($\alpha_{HR}$) at high-resolution (e.g., $\leq$0.001 cm$^{-1}$ for CO at 1 atm) using the methods described by Goldenstein et al. \cite{goldenstein2017spectraplot}. Here, the pertinent spectroscopic parameters were taken from the HITEMP 2019 database \cite{hargreaves2019hitemp}.  Gas properties (temperature ($T$), absorbing species column density ($\chi_{abs}L$), and, when relavent, pressure ($P$)) are free parameters in the fitting routine. H$_2$O was also included in the spectral simulations to account for weak interference within the interrogated wavelength band. Collisional-broadening parameters of CO lines by H$_2$O were taken from Hartmann et al. \cite{hartmann1988accurate}.
    
    \item Linearly interpolate an empirically determined spectrum of $I_0$ to the same resolution as $\alpha_{HR}$, yielding $I_{0,HR}$. Similarly, the high-resolution frequency axis is interpolated from the low-resolution frequency axis determined in step 2 on the measurement side.
    
    \item Generate a high-resolution spectrum of transmitted laser intensity ($I_{t,HR}$) with Beer's Law using $\alpha_{HR}$ and $I_{0,HR}$.
    
    \item Convolve $I_{t,HR}$ and $I_{0,HR}$ with the instrument response function to account for instrument broadening, yielding $I_{t,HR,conv}$ and $I_{0,HR,conv}$, respectively. In our prior work, parameters governing the shape of the IRF were floated in the fitting routine \cite{tancin2020ultrafast,tancin2021ultrafast,Radhakrishna2021, tancin2020ultrafast_AIAA}, however, here they were determined from the multi-spectrum fitting routine (discussed in Section \ref{section: multi_spect}) and held fixed during spectral-fitting.
    
    \item Interpolate $I_{t,HR,conv}$ and $I_{0,HR,conv}$ back to the lower spectral resolution of the data (yielding $I_{t,conv}$ and $I_{0,conv}$, respectively) using the binning interpolation method described in Section \ref{section: binning}.
    
    \item Use Beer's law to calculate the simulated spectral absorbance ($\alpha_{sim}$) from $I_{t,conv}$ and $I_{0,conv}$.
\end{enumerate}

Last, the spectra $\alpha_{sim}$ and $\alpha_{meas}$ are compared by the nonlinear least-squares fitting routine to determine if the fit has converged. If the convergence criteria has not been met, the procedure is repeated with new parameters until convergence.


\subsection{Binning interpolation}
\label{section: binning}

The high dispersion per pixel of ULAS measurements requires that absorbance spectra be simulated at a higher resolution than the measurement. Interpolating a spectrum between a high resolution simulation and the lower resolution measurement must be done in a manner which mirrors the physical discretization of the spectra by the camera's FPA.

\begin{figure}[ht!]
\centering
\includegraphics[width=1\textwidth]{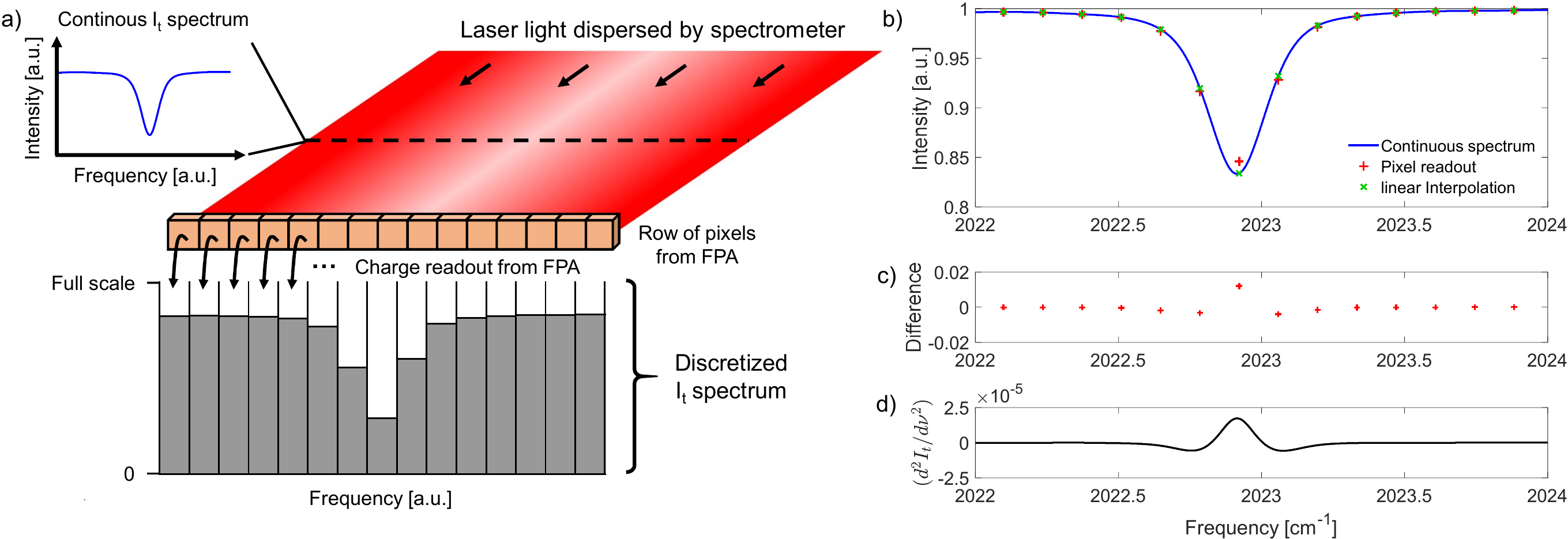}
\caption{A schematic illustrating the discretization process of an FPA, and the fidelity of binning and linear interpolation methods on mirroring this process. a) The discretization of a transmitted intensity spectrum by a row of pixels on the FPA. b) A plot comparing a continuous spectrum of $I_t$ with both a binning and linear down-sampling interpolation applied. c) Error (difference) between the binning and linear interpolation methods. d) Second derivative of transmitted intensity spectrum.}
\label{fig:interpolation}
\end{figure}

Figure \ref{fig:interpolation}a illustrates the discretization of a spectrum of $I_t$ by the camera's FPA. A laser pulse which has been dispersed and instrument broadened by a spectrograph approaches the camera's FPA with a continuous intensity spectrum. When the pulse hits the FPA, the light is collected by individual pixels, in effect, spatially binning the signal according to the pixel's physical dimensions. At the end of the camera's integration time, the camera reads out a single value for each pixel, which corresponds to the spatiotemporally averaged incident intensity. In the spectral-fitting routine, $I_{t,HR, conv}$ is the analog of the real-world continuous intensity spectrum. This spectrum must be down-sampled to the experimental frequency axis using the same binning method as the FPA, by calculating the average intensity across a given pixel.

The difference between ``binning interpolation'' and linear interpolation is illustrated in Fig. \ref{fig:interpolation}b-c. The error between interpolation methods is generally proportional to the second derivative of the spectrum being discretized, which is shown for reference in Fig. \ref{fig:interpolation}d. In essence, the differences between method of interpolation become significant when the intensity variation across a pixel cannot be assumed to be linear, because for linear variation, both a linear interpolation and the spatial mean value will be equal.

\section{Determination of instrument-broadening parameters}
\label{section: multi_spect}

\subsection{Background}
Accurate knowledge of line-broadening effects is crucial for correctly modeling measurements of absorption spectra, and hence, for accurate ULAS measurements of gas properties \cite{goldenstein2017infrared,hahn2010laser,park1994effect}. The broadening (or narrowing) of an absorption feature is due to a combination of physical (e.g., collisional broadening, Doppler broadening) and instrumental (e.g., the slit function of spectographs, spectrometers, or monochromators) broadening sources. Challenges arise when the FWHM of instrument-broadening sources are similar to, or greater than that of physical broadening sources, in which case error in one or the other can lead to inaccurate determination of gas properties.

Instrument-, Doppler-, and collisional-broadening have been the dominant sources of line-broadening in the ULAS measurements acquired to date \cite{tancin2020ultrafast, tancin2021ultrafast,Radhakrishna2021,tancin2020ultrafast_AIAA}. The Doppler broadening FWHM ($\Delta\nu_D$) has had the least uncertainty for given gas conditions, as it is only a function of molecular weight, temperature and linecenter frequency, which are either measured or relatively well known. The uncertainty in collisional-broadening FWHM ($\Delta\nu_C$) was much greater, due to uncertainty in both bath gas composition (e.g., in flames) and in collisional-broadening coefficients within databases. Often, this uncertainty is accounted for in spectral-fitting routines by floating either a scalar multiplier on all $\Delta\nu_C$ or floating $\Delta\nu_C$ directly for a sufficiently small number of lines \cite{goldenstein2017infrared}. Initially, instrument-broadening parameters were floated during the processing of ULAS measurements. However, simultaneously floating the IRF and collisional-broadening half-widths in the nonlinear fitting routine may be unsuitable due to the similar effect they have on transition lineshapes, and the high dispersion per data point (i.e., few pixels per absorption transition) which makes measurements insensitive to subtle differences in lineshapes. Potentially, the fitting routine may converge on incorrect values for both instrument and collisional-broadening parameters and result in substantial errors in measured gas properties. Therefore, it can be necessary to determine the IRF elsewhere so it can be held fixed in the spectral-fitting routine. In our experience, this is most critical when the measured absorbance spectrum does not possess a wide variety of linewidths, as is the case for the ULAS measurements in CO's P-branch at a single temperature or pressure.


There are numerous methods for determining the IRF of a spectrograph with a given optical setup. One of the most common is to measure the IRF through fitting a model to measurements of well known spectra \cite{lucht2003dual}. 
Another technique, is to measure the IRF directly by using a sufficiently narrowband (relative to the IRF FWHM) reference source such as a narrowband laser, low-pressure hollow-cathode lamp or a narrow, well-characterized absorption line \cite{hahn2010laser}. The IRF can be also be inferred in the case of a reference source with a non-negligible but well characterized linewidth \cite{heitmann1996measurements}. That said, these methods were not deemed suitable for high-accuracy ULAS measurements. As a result, a multi-spectrum fitting routine (MSFR) was used to determine instrument-broadening parameters.

\subsection{Multi-spectrum fitting routine}
\label{sect: MSFR}

In this work, the IRF was determined by applying an MSFR to ULAS data acquired at a variety of known gas conditions. The MSFR simultaneously fit multiple simulated absorbance spectra to absorbance spectra measured at multiple pressures while floating a single set of IRF parameters and a single correction factor for all $\Delta\nu_C$. The latter was done to account for uncertainty in the collisional-broadening parameters. Simultaneously fitting simulated spectra to data acquired at vastly different pressures allowed for more accurate and robust determination of instrument- and collisional-broadening parameters by decreasing the total number of free parameters per spectrum \cite{goldenstein2015diode}, and by decoupling the effects of instrument- and collisional-broadening. For example, spectra acquired at sufficiently low pressure are dominated by instrument broadening and are comparatively insensitive to errors in collisional-broadening parameters, while this effect is reversed at high pressure.


In the MSFR, the IRF was modeled by Eq. \ref{Eq: IRF} which represents the weighted sum of Gaussian and Lorentzian functions, with the FWHM and lineshape weighting (proportion of the lineshape which is Lorentzian vs. Gaussian) floated as free parameters.

\begin{equation}
\phi_{IRF}(\lambda) = w \phi_L(\lambda, FWHM)+(w-1)\phi_G(\lambda, FWHM)
\label{Eq: IRF}
\end{equation}

\begin{figure}[t!]
\centering
\includegraphics[width=0.85\textwidth]{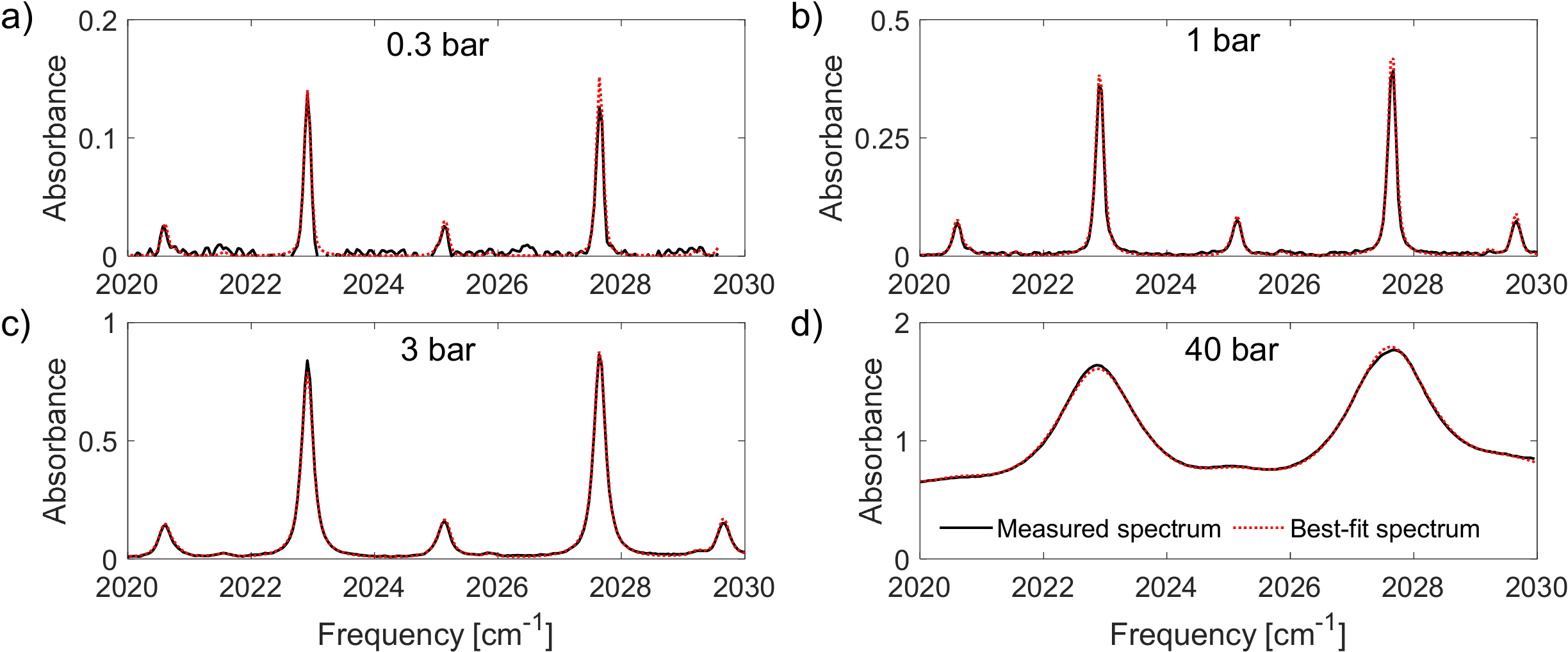}
\caption{Measured and best-fit absorbance spectra of CO acquired in a heated-gas cell at 1000 $\pm$ 20 K using a 0.3 nm resolution grating. Each best-fit spectrum was determined simultaneously using the MSFR. Spectra at (a) 0.3 bar, (b) 1 bar, (c) 3 bar, and (d) 40 bar are shown.}
\label{fig:multi_spectrum_spectra}
\end{figure}

\noindent Here, $\phi_G$ and $\phi_{L}$ are Gaussian and Lorentzian lineshapes, respectively, both with integrated areas of one, $w$ is the lineshape weighting parameter, $\lambda$ is the wavelength and $\phi_{IRF}$ is the IRF lineshape. Collisional-broadening coefficients for CO-CO$_2$ and CO-N$_2$ were taken from Hartmann et al. \cite{hartmann1988accurate}. A global scaling factor for $\Delta\nu_C$ was floated in the fitting routine as well. The frequency axis and a linear baseline correction were floated for each spectrum according to the data-processing procedure described in Section \ref{sect: fitting routine}, while temperature, pressure, and absorbing species concentration were held constant at their known values. Absorbance spectra measured at 0.3, 1, 3, and 40 bar within a heated-gas cell (described in Section \ref{sect: gas-cell setup}) were used for the MSFR. The residuals of each measured spectrum and best-fit were normalized by the peak absorbance to mitigate bias towards spectra with larger absorbance values. Figure \ref{fig:multi_spectrum_spectra} shows the measured and best-fit absorbance spectra determined by the MSFR for data acquired with the 300 lines/mm grating.

For data acquired with the 150 lines/mm grating, the MSFR returned a value of 0.520 nm for the IRF FWHM and an instrument lineshape which was 48.3\% Lorentzian. For the 300 lines/mm grating, the MSFR converged on an IRF FWHM of 0.297 nm and an instrument lineshape which was 61.5\% Lorentzian. These values compare relatively well to the gratings' theoretical resolutions (0.21 and 0.61 nm for the high and low resolution gratings, respectively), albeit, assuming a Gaussian IRF. The value for the scaling factor on $\Delta\nu_C$ was 0.91 for both gratings.

The accuracy of the parameters obtained by the MSFR were validated by processing data taken at each pressure individually according to the standard one-at-a-time data-processing procedure, with instrument-broadening parameters held fixed at the values determined by the MSFR. However, the factor on $\Delta\nu_C$ was floated for consistency with flame measurements where bath gas composition and pressure may not be known to high accuracy. Figure \ref{fig: hi-P gascell} illustrates the accuracy of the results. All measurements were either within 5\% of the known value, or within the bounds of the 95\% confidence intervals.

It is vital that the broadening parameters determined by the MSFR are valid outside of the dataset used in the MSFR. To test this, all heated-gas cell measurements were repeated 2 months later after the optical setup had been dismantled and rebuilt. The datasets were processed using the same instrument-broadening parameters as earlier (i.e., the MSFR was not rerun for these datasets). Gas properties were once again within 5\% of the known values, with the known value falling within the 95\% uncertainty bounds for most measurements.

For comparison, IRF parameters were also measured, using a quantum-cascade laser (QCL) as a reference source. The QCL emitted near 4.8 $\mu$m and had a negligible linewidth ($\approx$ 1 MHz) compared to the IRF of the spectrograph. The laser beam was focused into the spectrograph using the same optical setup as was used for ULAS measurements in heated-gas cell and propellant-flame tests. IRF parameters were then determined by fitting Eq. \ref{Eq: IRF} to the measured spectrum using a nonlinear fitting routine. The best-fit IRF FWHMs were 0.25 nm and 0.54 nm for the 300 lines/mm and 150 lines/mm gratings, respectively, which agree with values from the MSFR to within 4\% and 16\%. The slight disagreement compared to the MSFR can be explained by the low resolution of these QCL measurements (only 3-4 data points per spectrum) which make them less robust.

\subsection{Discussion of results}

The accuracy of the MSFR is supported by four key findings. (1) The broadening parameters determined through the MSFR enabled gas-cell validation measurements of temperature and absorbing species concentration which were within 5\% of the known values at pressures from 0.3 to 40 bar. (2) This accuracy was shown to be repeatable after rebuilding the optical setup. (3) The scaling factor on $\Delta\nu_C$ returned by the MSFR was the same for both of the grating resolutions which was, in turn, consistent with broadening uncertainties reported in the literature \cite{nair2020mhz}. (4) Last, the IRF FWHM agrees relatively well with values which were measured directly using a QCL. 


\section{Imaging ultrafast laser pulses in the mid-IR}
During the course of the development and initial applications of the ULAS diagnostic, several significant practical considerations for imaging ultrafast, mid-IR laser pulses were identified and are discussed below.

\subsection{Camera saturation}

The IR camera used for ULAS experiments (Telops FAST-IR 2K) \cite{tancin2020ultrafast, tancin2021ultrafast,Radhakrishna2021,tancin2020ultrafast_AIAA} has two modes of saturation which are relevant to ULAS measurements. The first is saturation of the photocurrent generation process in the FPA. If the signal flux (i.e., signal counts per unit time) is $\geq$ $\approx$ 5000 counts/$\mu$s (this value may range from 3000 to 6000 counts/$\mu$s for a given camera), additional incident photons will not be converted to signal counts. This is problematic in situations with short exposure times and high signal. For ULAS measurements, this was mitigated by using an integration time $\geq$ 4 $\mu$s, though, 5 $\mu$s was used to provide a margin of safety. 

The second relevant saturation mode corresponds to premature saturation of the number of possible signal counts. The typical response of the camera is between 0 and 65535 counts (i.e., 16-bit resolution), however, when imaging ultrafast pulses, saturation is observed at $\approx$ 18\% of its full range ($\approx$ 12,000 counts). It is currently hypothesized that this saturation mode is caused by depletion of charge carriers in the FPA and is distinct from the photocurrent saturation mode. This saturation mode is avoided by limiting signal counts to $\leq$ 10,000 counts, thus, allotting 2,000 counts of buffer for shot-to-shot laser fluctuations, drift in the time-averaged laser power, and emission during flame measurements.

Both of these saturation modes, even when avoided, have negative effects on ULAS measurements. Most notably, premature saturation limits the SNR of acquired spectra. Given that pixel noise is constant in magnitude, limiting signal counts to 12,000 reduces the potential SNR, and therefore detection limit, by a factor of five. Photocurrent saturation also limits SNR by limiting the camera's integration time and therefore the ability to reduce the contribution of the thermal background. For example, at an integration time of 5 $\mu$s the ambient thermal background typically contributes $\approx$ 1500 signal counts to acquired spectra, meaning that only 8,500 counts are left for recording laser intensity.

\subsection{Shot-to-shot repeatability}

\begin{figure}[t!]
\centering
\includegraphics[width=1\textwidth]{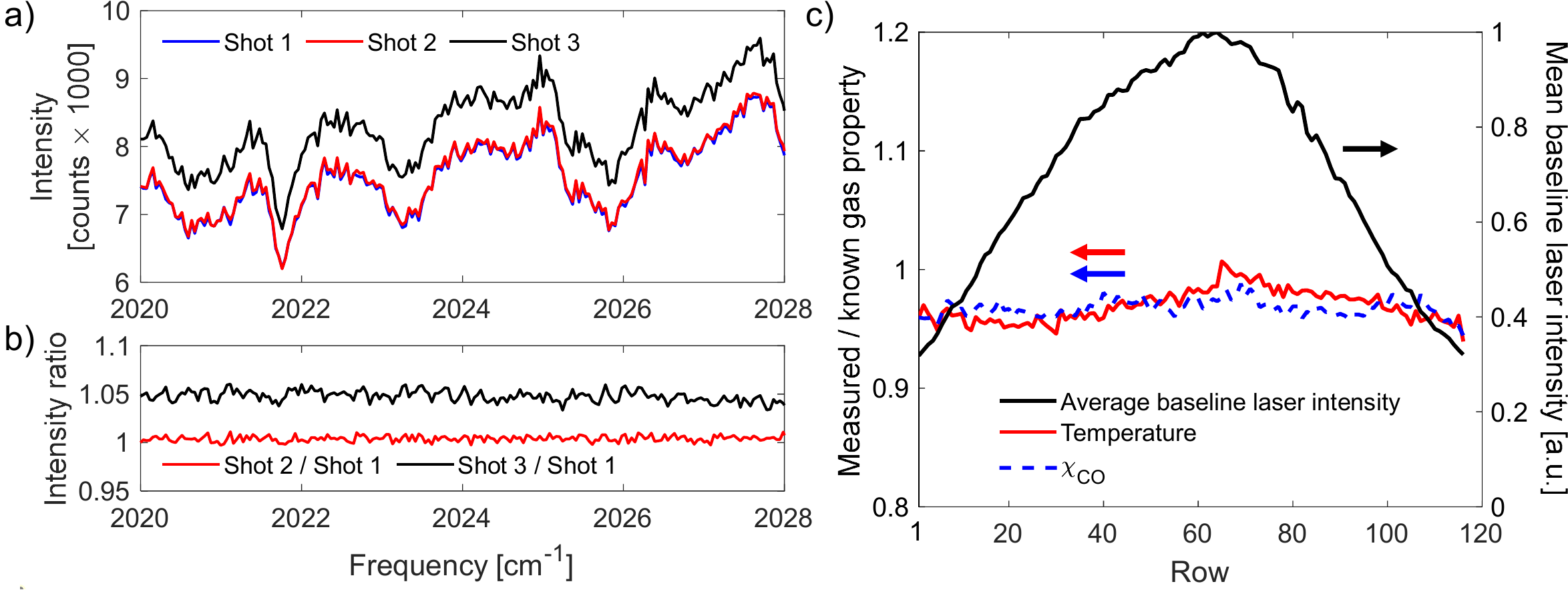}
\caption{(a) Three shots of $I_0$. Shot 1 and Shot 2 were recorded consecutively, and Shot 3 was recorded $\approx$ 6 minutes later. (b) Shots 2 and 3 are normalized to Shot 1 to highlight the stability in spectral shape, and suitability of a linear baseline correction. (c) The average value for $I_0$ is plotted as a function of FPA row, along with gas properties inferred from single-shot, single-row spectra to demonstrate the linear response of the IR camera to ultrafast laser pulses.}
\label{fig:linearity}
\end{figure}

Shot-to-shot repeatability, in both the spectral shape of the laser intensity, as well as the camera's response is vital for acquiring high-fidelity, single-shot ULAS measurements. While there are complex spectral variations in the baseline laser intensity (see Fig. \ref{fig:linearity}a), the shape of the spectrum is essentially constant with time, and mainly varies by a scalar multiple which enables ULAS to provide high-fidelity single-shot measurements. A small linear term was added to the scalar baseline correction which yielded further improvement. The stability of the baseline laser intensity between subsequent laser shots and over the course of $\approx$ 6 minutes is illustrated by Fig. \ref{fig:linearity}a-b.

\subsection{Pixel linearity}
The linearity of the InSb FPA was investigated to support the accuracy of ULAS measurements and assess the potential for 1D intra-pulse measurements of gas properties. Linearity was tested by processing and extracting gas properties from an image of transmitted intensity data which had pronounced intensity variation across its rows (the spatial dimension). A dataset from the gas cell experiments described in Section \ref{sect: MSFR} was selected since the gas conditions were spatially uniform. Data was acquired at 10 bar using the grating with 300 lines/mm and processed according to the methods described in Section \ref{sect: fitting routine}. The mean baseline laser intensity is shown in Fig. \ref{fig:linearity}c, which illustrates that the mean intensity varied by nearly a factor of three across rows of pixels. Gas properties inferred from single-shot, single-row spectra are also plotted as a function of row. The temperature and CO mole fraction varied by only 6\%, indicating that the camera's FPA responded linearly to the intensity of the ultrafast laser pulses. Further, this demonstrates the ability of ULAS to acquire 1D-resolved, single-shot measurements of gas properties.

\section{Experimental results and discussion}

\subsection{Validation tests}
\begin{figure}[t!]
\centering
\includegraphics[width=1\textwidth]{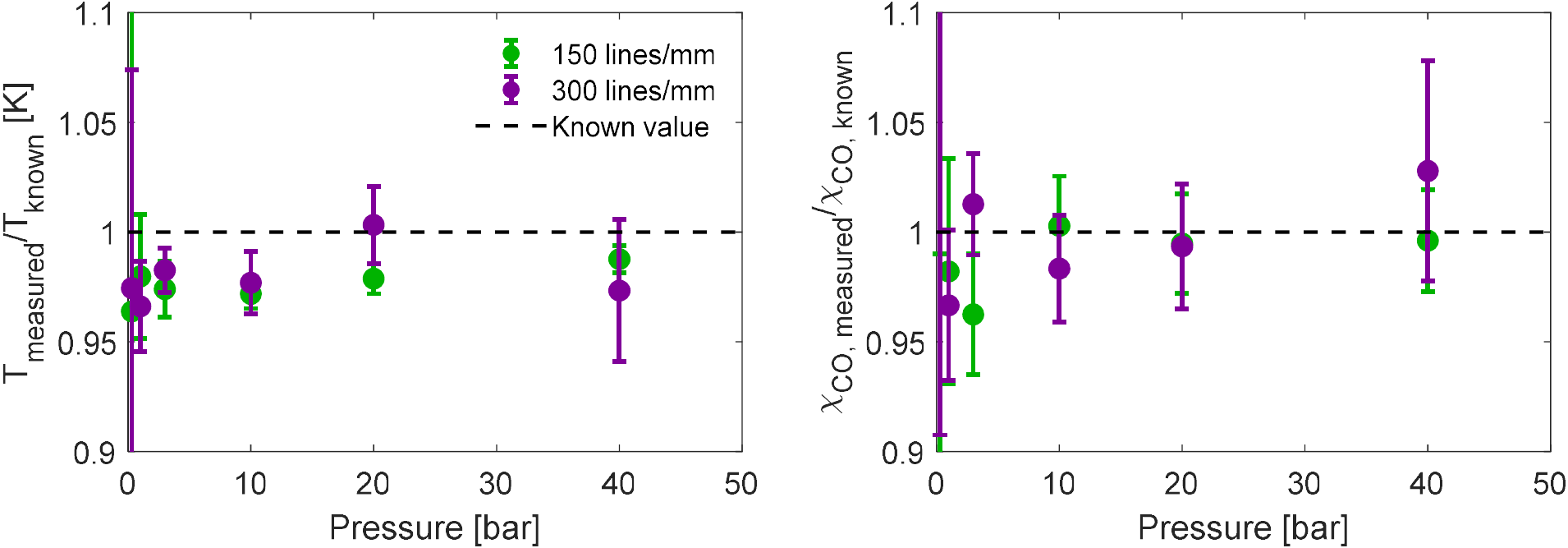}
\caption{Temperature and CO concentration measurements normalized to their known values acquired in a heated, static-gas cell at 1000 K. 
}
\label{fig: hi-P gascell}
\end{figure}

The accuracy of the ULAS diagnostic was validated at high temperature using the heated-gas cell data discussed in Section \ref{section: multi_spect}. Measurements of temperature and CO mole fraction were validated at 1000 $\pm$ 20 K and 0.3, 1, 3, 10, 20, and 40 bar. Instrument-broadening parameters were determined from the MSFR and held fixed for consistency with flame measurements. The results normalized to the known value are shown in Fig. \ref{fig: hi-P gascell}.

The ULAS measurements of temperature and CO mole fraction demonstrated high accuracy across a broad range of pressures, with values all falling within 5\% of the known value and most within the bounds of the 95\% confidence intervals. At 0.3 bar, the accuracy was lower compared to the high-pressure cases which was due to low signal levels.

\subsection{Characterization of AP-HTPB propellant flames}

\begin{figure}[ht!] 
\centering
\includegraphics[width=1\textwidth]{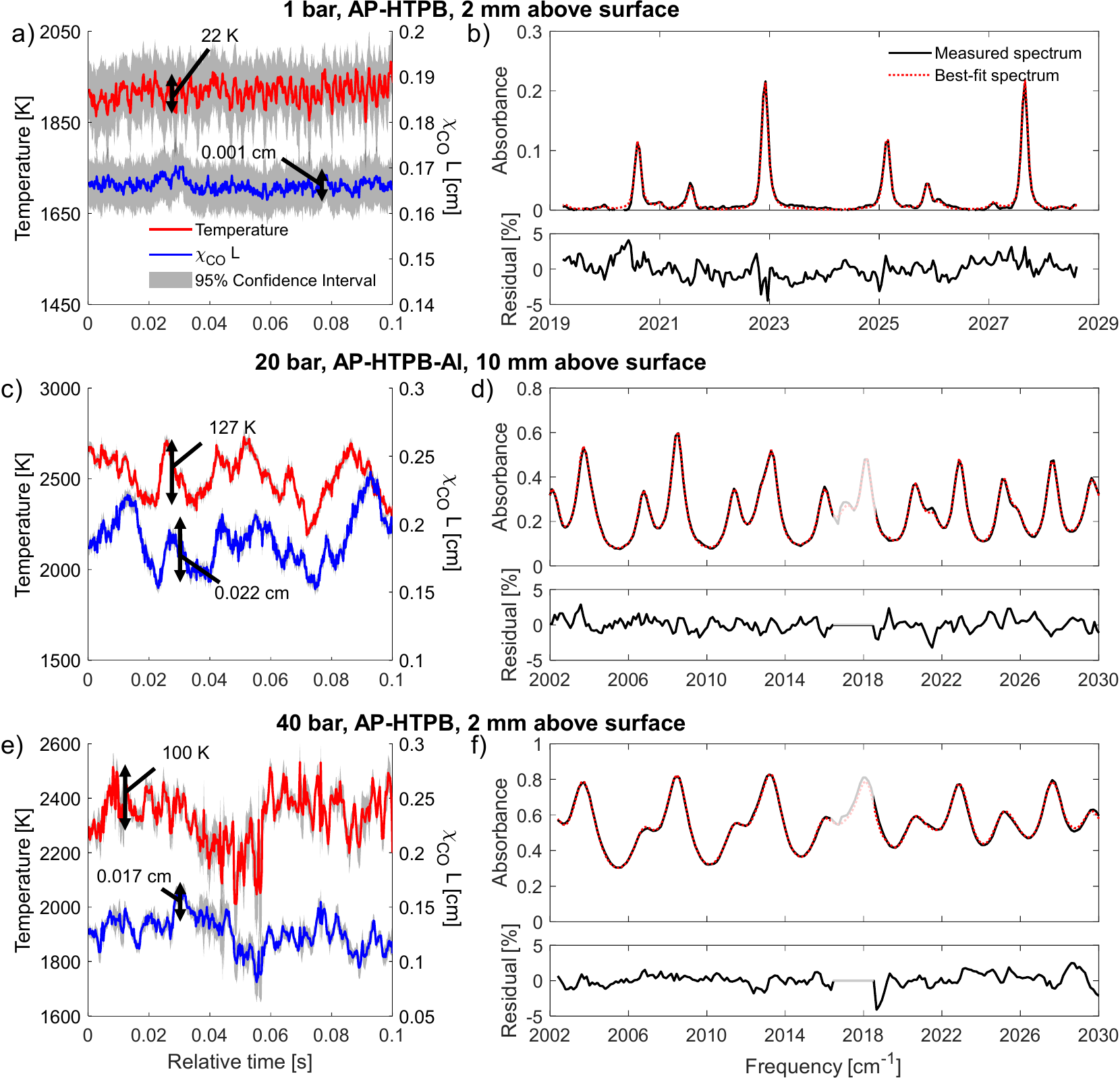}
\caption{Samples of measured time histories of temperature and $\chi_{CO}L$ (a, c, e) as well as representative measured and best-fit absorbance spectra of CO with residuals (b, d, f) acquired in propellant tests. The shaded out portion of the spectra shown in d) and f) was ignored by the spectral-fitting routine due interfering absorption by H$_2$O in the ambient air outside the combustion chamber. }
\label{fig:megaplot}
\end{figure}

Figure \ref{fig:megaplot} shows measured time histories of temperature ($T$) and CO column density ($\chi_{CO}L$) as well as representative examples of measured and best-fit absorbance spectra acquired in laser-ignited AP-HTPB flames at 1 and 40 bar and an AP-HTPB-Al flame at 20 bar. The spectra illustrate the high SNR of the measurements and, for all data sets, excellent agreement between the measured and simulated spectra, thereby supporting the accuracy of the spectroscopic model and fitting routine employed. The 1-$\sigma$ precision of the baseline noise level in absorbance was 0.003, 0.007 and 0.01 for the results shown in Fig. \ref{fig:megaplot}b, \ref{fig:megaplot}d and \ref{fig:megaplot}f, respectively. The time-averaged temperatures (1926 K at 1 bar and 2326 K at 40 bar) and CO mole fractions (0.2135 at 1 bar and 0.2138 at 40 bar) were slightly lower than constant enthalpy and pressure (HP) equilibrium calculations (i.e., 2417 K and 0.273 at 1 bar and 2483 K and 0.275 at 40 bar) \cite{ruesch2020characterization}. The lower than expected measured temperatures were likely due to a combination of heat loss, insufficient time and space for the product gas to fully reach equilibrium, and the path-integrated nature of the measurements \cite{tancin2020ultrafast_AIAA}. The lower than expected CO mole fractions were also likely affected by the product gas not fully reaching equilibrium (note the presence of large amounts of soot in Fig. \ref{fig:flame_pics}), as well as the substantial uncertainty in the absorbing path length. Notably, the measured CO mole fractions agree well with other LAS measurements in the literature \cite{ruesch2020characterization}. 

95\% confidence intervals for temperature and $\chi_{CO}L$ are plotted along with the measured time histories in Fig. \ref{fig:megaplot}a, c, and e. The corresponding 95\% confidence intervals for temperature and $\chi_{CO}L$ were 150 K and 0.01 cm for AP-HTPB at 1 bar, 61 K and 0.01 cm for AP-HTPB-Al at 20 bar, and 93 K and 0.02 cm for AP-HTPB at 40 bar. Larger uncertainty was observed during the 1 and 40 bar tests. At 1 bar, this is explained by the comparatively low signal, and narrow transition FWHMs (i.e., fewer data points per transition), which increased the uncertainty in fit, even with the higher-resolution grating being used. At 40 bar, non-absorbing transmission losses were the greatest of all tests due to severe beam steering, thereby, decreasing measurement SNR and increasing measurement uncertainty.

All of the time histories (see Fig. \ref{fig:megaplot}a,c,e) exhibit some unstructured variation in time, consistent with mild to moderate unsteadiness in general flame structure, which is supported through visual imaging of the flames (e.g., see Fig. \ref{fig:flame_pics}), particularly at high pressure. For example, measurements of temperature exhibited a 1-$\sigma$ variation of 22 K, 127 K, and 100 K at pressures of 1, 20, and 40 bar, respectively. Similarly, the 1-$\sigma$ variation in $\chi_{CO}L$ was 0.0012, 0.022, and 0.017 cm for pressures of 1, 20 and 40 bar, respectively. When accounting for the time-variation in gas properties (through subtracting the data from a 10-point moving average), the 1-$\sigma$ precisions at 1, 20 and 40 bar were 0.90\%, 0.91\%, and 1.6\%, respectively, of the measured value for temperature and 0.45\%, 2.0\%, and 3.7\% for $\chi_{CO}L$. Encouragingly, the results suggest that the addition of 15\% aluminum and, presumably, other metals into the propellant mixture does not significantly effect the performance of the diagnostic, thereby paving the way for future studies of metallized-propellant flames at rocket-motor-relevant pressures.

\section{Measurement uncertainty}

Uncertainty in measured gas properties was analyzed from three sources: (1) the instrument-broadening parameters, (2) noise in the measured spectra, and (3) the absorbing path length. Uncertainty in instrument-broadening parameters was quantified by calculating the 95\% confidence interval for each parameter using the covariance matrix returned by the MSFR. A Monte Carlo method was used to capture the complex and nonlinear relationship between the IRF parameters and inferred gas properties. Values for the IRF parameters were chosen as Gaussian random variables with the mean and standard deviation determined from the 95\% confidence intervals. The spectral-fitting routine was then run for 100 trials at each pressure and grating resolution, resulting in distributions of inferred gas properties which were determined to be normal by the Shapiro-Wilk test with a p-value of 0.05. 95\% confidence intervals from uncertainty in instrument broadening were thus determined from these distributions. For flame measurements, this was done for a single spectrum chosen soon after the flame reached steady state, and the value for uncertainty was assumed constant for the rest of the spectra acquired during steady-state burning. Uncertainty in inferred gas properties due to noise in the measured spectra for the one-at-a-time fitting routine were similarly calculated from the covariance matrix. Uncertainty in the absorbing path-length was set at 2\% for gas cell experiments based on the perceived error in the length of the CaF$_2$ rods, uncertainty in the thickness of the epoxy seal for the rods, and the uncertainty in thermal expansion of the gas cell and CaF$_2$ rods at high temperatures \cite{schwarm2019high}. For flame experiments, the uncertainty in absorbing path length was determined from simultaneous visual imaging, and estimated to be 10\% of the measurement. The uncertainty from all sources were added in quadrature to yield the total uncertainty for the gas properties. The calculated 95\% confidence intervals are shown in Fig. \ref{fig: hi-P gascell} and Fig. \ref{fig:megaplot}.

\section{Conclusions}
This work presented the extension of ULAS to calibration-free, single-shot measurements of temperature and CO across a broad range of pressures. Numerous improvements in the optical setup and the data-processing routine made this possible. An MSFR was utilized to accurately determine IRF parameters which enabled rigorous validation of the diagnostic across a wide range of pressures. AP-HTPB flames both with and without aluminum were characterized at pressures up to 40 bar. Further, the proof-of-concept for 1D-resolved measurements with sub-nanosecond time resolution was presented. Collectively, these results demonstrate that ULAS can be used for high-fidelity characterization of multi-phase combustion gases with improved temporal resolution at rocket-motor relevant pressures.

\section*{Funding}
The authors would like to thank the Air Force Office of Scientific Research (Grant: FA9550-18-1-0210) with Dr. Mitat Birkan as program manager and the National Science Foundation CBET Award No. 1834972 for supporting this work.

\section*{Acknowledgments}
The authors would like to thank Morgan Ruesch and Prof. Steven Son for providing and preparing the propellant used for this work, as well as Prof. Robert P. Lucht, Vishnu Radhakrishna, and Ziqiao Chang for helping support the operation and maintenance of the ultrafast-laser system.

\bibliographystyle{style}
\bibliography{references}

\end{document}